\documentclass
[showpacs,superscriptaddress,prl,twocolumn,tightenlines,balancelastpage,10pt,a4paper]{revtex4}%
\usepackage{amsmath}
\usepackage{amsfonts}
\usepackage{amssymb}
\usepackage{graphicx}%
\begin{document}
\title{Experimental Quantum Error Rejection for Quantum Communication}
\author{Yu-Ao Chen}
\affiliation{Physikalisches Institut, Universit\"{a}t Heidelberg,
Philosophenweg 12, D-69120 Heidelberg, Germany}
\author{An-Ning Zhang}
\affiliation{Hefei National Laboratory for Physical Sciences at
Microscale and Department of Modern Physics, University of Science
and Technology of China, Hefei, Anhui 230027, People's Republic of
China}
\author{Zhi Zhao}
\affiliation{Physikalisches Institut, Universit\"{a}t Heidelberg,
Philosophenweg 12, D-69120 Heidelberg, Germany} \affiliation{Hefei
National Laboratory for Physical Sciences at Microscale and
Department of Modern Physics, University of Science and Technology
of China, Hefei, Anhui 230027, People's Republic of China}
\author{Xiao-Qi Zhou}
\affiliation{Hefei National Laboratory for Physical Sciences at
Microscale and Department of Modern Physics, University of Science
and Technology of China, Hefei, Anhui 230027, People's Republic of
China}
\author{Jian-Wei Pan}
\affiliation{Physikalisches Institut, Universit\"{a}t Heidelberg,
Philosophenweg 12, D-69120 Heidelberg, Germany}\affiliation{Hefei
National Laboratory for Physical Sciences at Microscale and
Department of Modern Physics, University of Science and Technology
of China, Hefei, Anhui 230027, People's Republic of China}

\date{\today }

\begin{abstract}
We report an experimental demonstration of a bit-flip error
rejection protocol for error-reduced transfer of quantum
information through a noisy quantum channel. In the experiment, an
unknown state to be transmitted is encoded into a two-photon
entangled state, which is then sent through an engineered noisy
quantum channel. At the final stage, the unknown state is decoded
by a parity measurement, successfully rejecting the erroneous
transmission over the noisy quantum channel.

\end{abstract}
\maketitle

A crucial step in the full realization of long-distance quantum
communication is to overcome the problems caused by decoherence
and exponential photon loss in the noisy quantum channel
\cite{gisinrmp}. As a general solution, two distant parties could
first share highly entangled photon pairs, the transmission of
quantum states for various applications in quantum communication
can then be achieved by using ancilla entanglement. As the quantum
repeater \cite{Briegel98}, combining entanglement purification
\cite{bennett96} and entanglement swapping \cite{zuk93}, could
provide an efficient way to generate highly entangled states
between distant locations, many experimental efforts have been
made to achieve entanglement swapping, entanglement purification
and quantum memory \cite{pan98a,pan03,kuzmich04,polzik04}, and
even the demonstration of a prototype of quantum relay
\cite{zhao03,gisin04}. However, one still has a long way to go
before the above techniques can be realistically applied to
long-distance quantum communication.

Meanwhile, in the context of quantum error correction (QEC) the
way to protect a fragile unknown quantum state is to encode the
state into a multi-particle entangled state
\cite{shor95,steane96,laflamme96}. Then, the subsequent
measurements, i.e. the so-called decoding processes, can find out
and correct the error during the quantum operations. Several QEC
protocols have been experimentally demonstrated in the NMR
\cite{cory,knill} and ion-trap \cite{ion} systems. Although the
QEC are primarily designed for large scale quantum computing, the
similar idea was also inspired to implement error-free transfer of
quantum information through a noisy quantum channel \cite{dik01}.

The main idea in the original scheme is to encode an unknown
quantum state of single particle into a two-particle entangled
state \cite{dik01}. After the encoded state is transmitted over
the noisy quantum channel, a parity check measurement
\cite{pan98b} is sufficient to reject the transmission with
bit-flip error. Such a scheme has the advantage of avoiding the
difficult photon-photon controlled-NOT gates necessary for the
usual QEC. Moreover, the error rejection scheme proposed promises
additional benefit of high efficiency, compared to the QEC based
on linear optics quantum logic operations \cite{klm01}. This is
because the crucial feed-forward operations in linear optics QEC
will lead to very low efficiency. Although the original scheme is
within the reach of the current technology as developed in the
recent five-photon experiments \cite{zhao04,zhao05}, it is not
optimal in its use of ancilla entangled state because the encoding
process is implemented via a Bell-state measurement.

Remarkably, it is found recently \cite{wang04} that one pair of
ancilla entangled state is sufficient to implement the two-photon
coding through two quantum parity measurements. Thus, an elegant
modification of the previous experiment on four-photon
entanglement \cite{pan01-1} would allow a full experimental
realization of the error rejection code.

In this letter, we report an experimental realization of bit-flip
error rejection for fault-tolerant transfer of quantum states
through a noisy quantum channel. An unknown state to be
transmitted is first encoded into a two-photon entangled state,
which is then sent through an engineered noisy quantum channel. At
the final stage, the unknown state is decoded by quantum parity
measurement, successfully rejecting the erroneous transmission
over the noisy quantum channel.

\begin{figure}
[ptb]
\begin{center}
\includegraphics[
height=2.0092in, width=3.2846in] {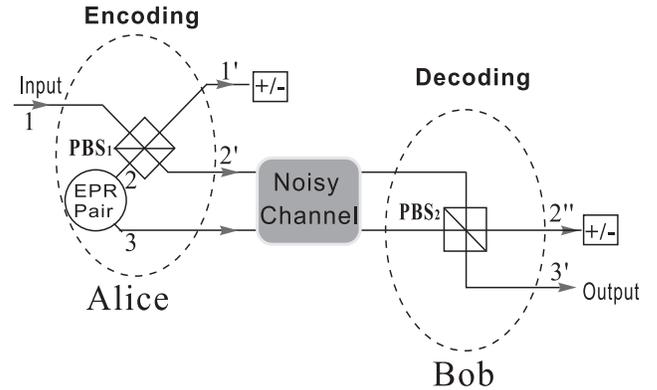} \caption{Scheme for
one bit-flip quantum error-rejection.} \label{figure1}
\end{center}
\end{figure}

Let us first consider the scenario that Alice wants to send a
single photon in an unknown polarization state $\alpha\left\vert
H\right\rangle_1 +\beta \left\vert V\right\rangle_1$ to Bob
through a noisy quantum channel. As shown in Fig. 1, instead of
directly sending it to Bob, Alice can encode her unknown state
onto a two-photon entangled state with an ancilla pair of
entangled photons:
\begin{equation}
\left\vert \phi^{+}\right\rangle _{23}=\tfrac{1}{\sqrt{2}}\left(
\left\vert HH\right\rangle _{23}+\left\vert VV\right\rangle
_{23}\right)  .
\end{equation}
The photon in the unknown polarization state and one photon out of
the ancilla entangled photon\ are superposed at a polarization
beam splitting (PBS$_{1}$). Behind the PBS$_{1}$, with a
probability of 50\% we obtain the renormalized state corresponding
to the three-fold coincidence among modes $1^\prime$, $2^\prime$
and 3
\begin{equation}
\left\vert \psi\right\rangle
_{1^{\prime}2^{\prime}3}=\alpha\left\vert HHH\right\rangle
_{1^{\prime}2^{\prime}3}+\beta\left\vert VVV\right\rangle
_{1^{\prime}2^{\prime}3}.
\end{equation}
Conditional on detecting photon $1^{\prime}$ in the $\left\vert
+\right\rangle $ polarization state with a probability of 50\%,
where $\left\vert \pm\right\rangle =\frac{1}{\sqrt{2}}\left(
\left\vert H\right\rangle \pm\left\vert V\right\rangle \right)$,
the remaining two photons will then be projected onto the
following entangled state:
\begin{equation}
\left\vert \psi\right\rangle _{2^{\prime}3}=\alpha\left\vert HH\right\rangle
_{2^{\prime}3}+\beta\left\vert VV\right\rangle _{2^{\prime}3}.
\end{equation}
Thus, through a quantum parity measurement between modes
$1^\prime$ and $2^\prime$, a two-photon encoding operation can be
realized.

After finishing the encoding process, Alice sends photons
$2^{\prime}$ and 3 to Bob through a noisy quantum channel and Bob
will recombine the two photons at the PBS$_{2}$ in order to
identify and reject the erroneous transmission. If there is no
error in the quantum channel, Bob will obtain the same quantum
state as in (3) after PBS$_{2}$. Projecting photon
$2^{\prime\prime}$ into the $\left\vert +\right\rangle $ state
with a success probability of $50\%$, photon $3^{\prime}$ will be
left in the unknown state $\alpha\left\vert H\right\rangle
+\beta\left\vert V\right\rangle $. Through the decoding process,
i.e. conditional on detecting in mode $2^{\prime\prime}$ one and
only one $\left\vert +\right\rangle $-polarized photon, Bob can
recover the state originally sent by Alice.

If a bit-flip error occurred for one of the two transmitted
photons, the two photons will have different polarizations and
exit the PBS$_{2}$ in the same output arm. Therefore, no
coincidence will be observed between modes $2^{\prime\prime}$ and
$3^\prime$. That is to say, the bit-flip error during the
transmission of quantum states over the noisy channel has been
simply rejected by the final quantum parity measurement. However,
if both bit-flip errors occurred simultaneously for the two
transmitted photons, Bob would finally obtain the polarization
state of $\alpha\left\vert V\right\rangle +\beta\left\vert
H\right\rangle $ via the same quantum parity measurement for
decoding operation and the error can not be effectively rejected.

Moreover, the detection of photon $1'$ in the
$\left\vert-\right\rangle$ state also leads to encoding of the
initial quantum state in a two photon state, provided the
associated phase flip is taken into account. Obviously, the same
holds for the decoding at Bob's: projection onto the
$\left\vert-\right\rangle$ state is associated with a phase flip
that can be compensated for. The coding and decoding efficiency
can thus be increased by a factor of two each.

Specifically, suppose that Alice would send photons to Bob in one
of the three complementary bases of $\left\vert
H\right\rangle/\left\vert V\right\rangle $, $\left\vert
+\right\rangle/\left\vert -\right\rangle $ and $\left\vert
R\right\rangle/\left\vert L\right\rangle $ (where $\left\vert
R\right\rangle =\tfrac{1}{\sqrt{2}}\left( \left\vert
H\right\rangle + i\left\vert V\right\rangle \right)$, and
$\left\vert L\right\rangle =\tfrac{1}{\sqrt{2}}\left( \left\vert
H\right\rangle - i\left\vert V\right\rangle \right)$), and each
qubit is directly sent through the noisy quantum channel with
bit-flip error rate of $E_{0}=p$. The quantum bit error rate
(QBER) after the decoding process \cite{wang04} will be:
\begin{equation}
E_{1}=\frac{p^{2}}{\left(1-p\right)^{2}+p^{2}},
\end{equation}
for the polarization states of $\left\vert
H\right\rangle/\left\vert V\right\rangle $ and $\left\vert
R\right\rangle/\left\vert L\right\rangle $, and no error occurs
for the $\left\vert +\right\rangle/\left\vert-\right\rangle $.
Therefore, the QBER of $E_{1}$ will be lower, compared to the QBER
of $E_{0}$ for any $p<1/2$. For small $p$, $E_{1}$ is on the order
of $p^{2}$. The transmission fidelity can thus be greatly improved
by using the quantum error rejection code.

Note that, conditional detection of photons in mode $1^{\prime}$
implies that there is either zero or one photon in the mode
$2^{\prime}$. But, as any further practical application of such a
coding involves a final verification step, detecting a threefold
coincidence makes sure that there will be exactly one photon in
each of the modes $2^{\prime} $ and $3$. This feature allows us to
perform various operations like, for example, the recombination of
two photons at PBS$_{2}$ before the final detection. This makes
our encoding scheme significantly different from a previous
two-photon encoding experiment \cite{pittman04}, where there are
certain probabilities of containing two photons in one of two
encoding modes. Thus, the previous two-photon encoding experiment
cannot be applied to the error-rejection code.

Moreover, we would like to emphasize that, compared to the two
recent experiments on fault-tolerant quantum information
transmission \cite{Ricci04,Massar05} our protocol has two
essential advantages. On the one hand, the work in \cite{Ricci04}
can only encode and send a known state instead of encoding and
sending arbitrary unknown states required by many quantum
communication protocols. On the other hand, the experiment in
\cite{Massar05} can only filtrate half of the single phase-shift
error. Thus, if the error rate of the channel is $p$, after
applying the error filtration method the remaining QBER is still
larger than $p/2$ even in the ideal case. Note that, the error
filtration probability in \cite{Massar05} can be increased by
coding a qubit in a larger number of time-bins, however, this
would need much more resources. While our method can in principle
reject any one bit-flip error with certainty as analyzed before.
In fact, the ability to suppress the first order error ($p$) to
the second order ($p^2$) is essential to overcome the channel
noise in scalable quantum communication.

\begin{figure}
[ptb]
\begin{center}
\includegraphics[
height=2.5097in] {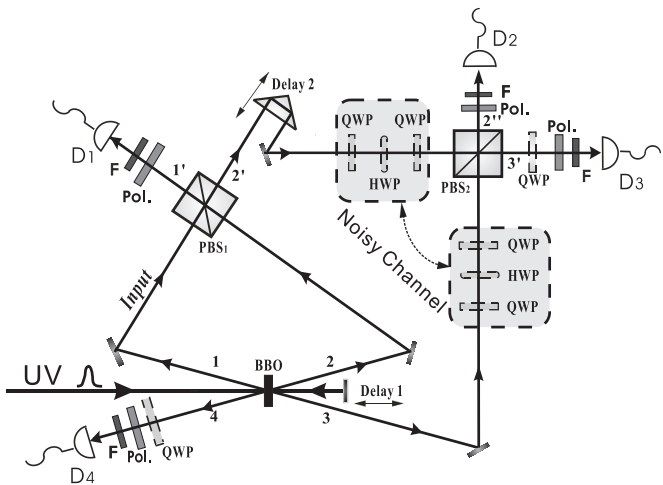} \caption{Experimental set-up for
quantum error-rejection.} \label{figure2}
\end{center}
\end{figure}

A schematic drawing of the experimental realization of the error
rejection is shown in Fig. 2. A UV pulse (with a duration of
200fs, a repetition rate of 76MHz and a central wavelength of
394nm) passes through a BBO crystal twice to generate two
entangled photon pairs $1 $, $4$ and $2$, $3$ in the state
$\left\vert \phi^{+}\right\rangle $ \cite{kwiat}. The high quality
of two-photon entanglement is confirmed by observing a visibility
of $(94\pm1)\%$ in the
$\left\vert+\right\rangle/\left\vert-\right\rangle$ basis. One
quarter wave plate (QWP) and one polarizer (Pol.) in front of
detector $D_4$ are used to perform the polarization projection
measurement such that the input photon in mode 1 is prepared in
the unknown state.

The two photons in modes 1 and 2 are steered to the PBS$_{1}$
where the path length of photon $1$\ have been adjusted by moving
the delay mirror Delay 1 such that they arrive simultaneously.
Conditional on detecting photon $1^{\prime}$ in the $\left\vert
+\right\rangle $ polarization, the unknown polarization state was
encoded into the modes in $2^{\prime}$ and $3$. The encoded
two-photon state is transmitted through the engineered noisy
quantum channel and then recombined at the PBS$_{2}$. Furthermore,
the path length of photon $2^{\prime}$ has been adjusted by moving
the Delay 2 such that photons in modes $2^{\prime}$ and $3$ arrive
at the PBS$_{2}$ simultaneously. Through the whole experiment,
spectral filtering (with a FWHM 3nm, F in Fig. 2) and
fiber-coupled single-photon detectors have been used to ensure
good spatial and temporal overlap between photons in modes $1$ and
$2$, and photons in modes $2^{\prime}$ and $3$ \cite{zuk01}.

To characterize the quality of the encoding and decoding process,
we first measure the interference visibility at the PBS$_1$. Since
photon pairs 1-4 and 2-3 are in the state
$\left\vert\phi^{+}\right\rangle$, it is easy to see that the
four-fold coincidence in $1^{\prime}$, $2^{\prime}$, $3$ and $4$
corresponds to a four-photon GHZ state
$\tfrac{1}{\sqrt{2}}(\left\vert HHHH\right\rangle
_{1^{\prime}2^{\prime}34}+\left\vert VVVV\right\rangle
_{1^{\prime}2^{\prime}34})$ \cite{pan01-1}. The four-photon
entanglement visibility in the
$\left\vert+\right\rangle/\left\vert-\right\rangle$ basis was
observed to be $(83\pm3)\%$. Similarly, the four-photon
entanglement visibility in modes $1^{\prime}$, $2^{\prime\prime}$,
$3^{\prime}$ and $4$ is observed to be $(80\pm3)\%$, before
introducing artificial channel noise. Note that, the visibility is
obtained after compensating the birefringence effect of the PBSes
\cite{pan03-1}.

In the experiment, the noisy quantum channels are simulated by one
half wave plate (HWP) sandwiched with two QWPs. Each of two QWP is
set at 90$^{0}$ such that the horizontal and vertical polarization
will experience 90$^{0}$ phase shift after passing through the
QWPs. By randomly setting the HWP axis to be oriented at
$\pm\theta$ with respect to the horizontal direction, the noisy
quantum channel can be engineered with a bit-flip error
probability of $p=\sin^{2}\left( 2\theta\right)$.

In order to show that our experiment has successfully achieved the
error rejection code, the quantum states to be transmitted in mode
$1$ are prepared along one of the three complementary bases of
$\left\vert H\right\rangle/\left\vert V\right\rangle$, $\left\vert
+\right\rangle/\left\vert-\right\rangle$, and $\left\vert
R\right\rangle/\left\vert L\right\rangle$. The error rates in the
engineered quantum channel can be varied by simultaneously
changing the axis of each half-wave plate. Specifically, we vary
the angle $\theta$ to achieve various error rates from 0 to $0.40$
with an increment $0.05$ in the quantum channel.

\begin{figure}
[ptb]
\begin{center}
\includegraphics[
width=3.5887in] {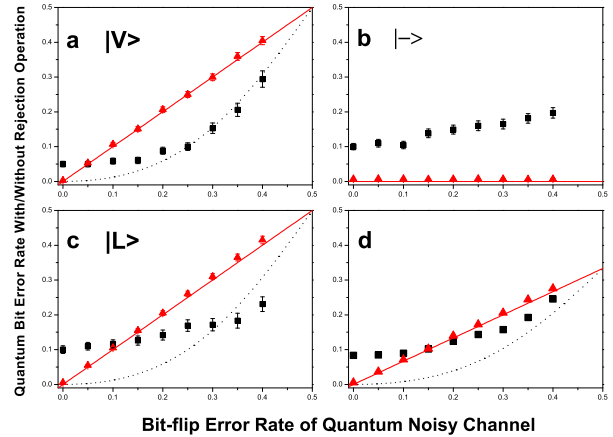} \caption{Experimental results for
three different initial states $\left\vert V\right\rangle$ (a),
$\left\vert -\right\rangle$ (b) and $\left\vert L\right\rangle$
(c), and (d) shows the average QBER (calculated over all the six
states). The quadrangle and the triangle dots are corresponding to
the cases that error-rejection and no error-rejection was made,
and the solid curves and the dot curves show the theoretically
prediction of QBER for the cases without and with error rejection
respectively.}
\end{center}
\end{figure}

The experimental results of three different input states, after
passing through the noisy quantum channel, are shown in Fig. 3,
The triangle dots in Fig. 3, corresponding to the bit-flip error
rates of single photons, were measured by directly sending the
quantum state of photon 1 (after passing through a PBS and some
wave-plates for state preparing) through the engineered quantum
channel while with both PBS$_1$ and PBS$_2$ removed. These dots
also shows the quality of the simulated error of the quantum noisy
channel. The quadrangle dots show the final bit-flip error rates
after performing error rejection operation with the help of
PBS$_1$ and PBS$_2$. Fig. 3a, 3b and 3c shows the experimental
results for the input states $\left\vert V\right\rangle$,
$\left\vert -\right\rangle$, and $\left\vert L\right\rangle$,
respectively. The other three input states have the similar
results as the one with the same basis respectively. And Fig. 3d
shows the average QBER calculated over all six input states.

In Fig. 3, one can clearly see that our error-rejection operation
itself also introduces  significant error rates, even with
$E_0=0$. Therefore, if the original $E_0$ is comparable with the
error rate caused by the experimental imperfection, no improvement
will be gained after error-rejection. In the $\left\vert
H\right\rangle$/$\left\vert V\right\rangle$ experiment, the
experimental error rate is about $5\%$. In both $\left\vert
+\right\rangle$/$\left\vert -\right\rangle$ and $\left\vert
R\right\rangle$/$\left\vert L\right\rangle$ experiments an
experimental error rate of $10\%$ is observed.

We notice that, whereas both the $\left\vert
+\right\rangle$/$\left\vert -\right\rangle$ and $\left\vert
R\right\rangle$/$\left\vert L\right\rangle$ experiments have
roughly the same visibility, a better visibility is obtained in
the $\left\vert H\right\rangle$/$\left\vert V\right\rangle$
experiment. This is mainly due to our two-photon entanglement
source, which has a better visibility in the $\left\vert
H\right\rangle$/$\left\vert V\right\rangle$ basis ($97\%$) than in
the $\left\vert +\right\rangle$/$\left\vert -\right\rangle$ or
$\left\vert R\right\rangle$/$\left\vert L\right\rangle$ basis
($94\%$). Moreover, it is partly due to the imperfect birefringent
compensation at the PBS$_1$ and PBS$_2$ \cite{pan03-1}, which
leads a reduction of interference visibility, hence imperfect
encoding and decoding process. Moreover, the imperfect encoded
state passing through the noisy channel also leads that in the
$\left\vert +\right\rangle$/$\left\vert -\right\rangle$ basis the
result becomes deteriorate as increasing of artificial noise.

From Fig. 3a and 3c, it is obvious that our error-rejection method
can significantly reduce the bit-flip error as long as $E_0$ is
larger than the experimental error rates. However, although
ideally in the $\left\vert +\right\rangle $/$\left\vert
-\right\rangle$ experiment no error should occur after the
error-rejection operation, an error rate no less than $10\%$ is
observed, which is in accordance with the limited visibility of
$80\%$.

Although our experimental results are imperfect, they are
sufficient to show a proof of principle of a bit-flip error
rejection protocol for error-reduced transfer of quantum
information through a noisy quantum channel. Moreover, Fig. 3d
shows that for a substantial region our experimental method does
provide an improved QBER over the standard scheme in a six-state
quantum key distribution (QKD). This implies, with further
improvement, the error-rejection protocol may be used to improve
the threshold of tolerable error rate over the quantum noisy
channel in QKD \cite{wang01}.

Our experimental realization of bit-flip error rejection deserves
some further comments. First, the same method can be applied to
reject the phase-shift error because phase errors can be
transformed into bit-flip errors by a $45^0$ polarization
rotation. In this way we can reject all the 1 bit phase-shift
error instead of bit-flip error. Second, by encoding unknown
states into higher multi-photon ($N$-photon) entanglement and
performing multi-particle parity check measurement \cite{pan98b}
either the higher order (up to $N-1$) bit-flip error or
phase-shift error can be rejected for more delicate quantum
communication.

In summary, our experiment shows a proof of principle of a
bit-flip error rejection protocol for error-reduced transfer of
quantum information through a noisy quantum channel. Moreover, by
further improvement of the quality of the resource for
multi-photon entanglement, the method may also be used to enhance
the bit error rate tolerance \cite{Chau,lo} over the noisy quantum
channel and offer a novel way to achieve long-distance
transmission of the fragile quantum states in the future QKD.

\begin{acknowledgments}
We are grateful to X.-B. Wang for valuable discussion. This work
was supported by the Alexander von Humboldt Foundation, the Marie
Curie Excellence Grant of the EU and the Deutsche Telekom
Stiftung. This work was also supported by the NNSFC and the CAS.
\end{acknowledgments}

\end{document}